\documentclass[journal]{IEEEtran}
\usepackage{amsmath}  % From the American Mathematical Society
\usepackage{amssymb}

  \usepackage{graphicx}
\usepackage{cite} % useful option \usepackage[noadjust]{cite} 

\begin{document}

\title{High Speed Chaos in Optical Feedback System with Flexible Timescales}

%\author{J. N. Blakely}
%\affiliation{Department of Physics and Center for Nonlinear and Complex Systems, Duke University, P.O. Box 90305, Durham, NC 27708}
%
%
%\author{L. Illing}
% \email{illing@phy.duke.edu}
%\affiliation{Department of Physics and Center for Nonlinear and Complex Systems, Duke University, P.O. Box 90305, Durham, NC 27708}
%
%
%\author{Daniel J. Gauthier}
%\affiliation{Department of Physics and Center for Nonlinear and Complex Systems, Duke University, P.O. Box 90305, Durham, NC 27708}
%
%
% here is the IEEE style for authors
%
\author{%
J. N. Blakely, Lucas Illing and Daniel J. Gauthier%
\thanks{%Manuscript received ??, 2003; revised ??, 2003.%
This work was supported by the US Army Research Office (grant \# DAAD19-99-1-0199).}% 
\thanks{ J. N. Blakely is with the US Army Research, Development, and Engineering Command, AMSRD-AMR-WS-ST, Redstone Arsenal, Alabama 35898}% 
\thanks{Lucas Illing and Daniel J. Gauthier are with the Department of Physics and the Center for Nonlinear and Complex Systems, Duke University, P.O. Box 90305, Durham, NC 27708 (e-mail:illing@phy.duke.edu)}%
}

%\date{\today}

% make the title area
\maketitle

\begin{abstract}
We describe a new opto-electronic device with time-delayed feedback that uses a Mach-Zehnder interferometer as passive nonlinearity and a semiconductor laser as a current-to-optical-frequency converter. Bandlimited feedback allows tuning of the characteristic time scales of both the periodic and high dimensional chaotic oscillations that can be generated with the device. Our implementation of the device produces oscillations in the frequency range of tens to hundreds of MHz. We develop a model and use it to explore the experimentally observed Andronov-Hopf bifurcation of the steady state and to estimate the dimension of the chaotic attractor.
\end{abstract}

\begin{keywords}
Feedback lasers, Optoelectronic devices, Electrooptic devices, Delay effects, Bifurcation, Chaos, Nonlinear systems, Nonlinear differential equations
\end{keywords}

% For peerreview papers, inserts a page break and creates the second title.
% Will be ignored for other modes.
\IEEEpeerreviewmaketitle

\section{Introduction}

\PARstart{T}{ime} 
delay systems are widely used as generators of chaos in applications such as chaos communication \cite{greg&roy-science,goedgebuer-prl,greg&roy-prl1,Davis_pra,Ohtsubo_opt_lett,greg&roy-prl2,illing_prl} and chaos control \cite{Pyragas,Meucci,Gavrielides,Gauthier&Sukow,Gavrielides&Sukow}. Many of the experiments conducted so far employed lasers with delayed feedback~\cite{greg&roy-science,goedgebuer-prl,greg&roy-prl1,Davis_pra,Ohtsubo_opt_lett,greg&roy-prl2,Pyragas,Meucci,Gavrielides,Gauthier&Sukow,Gavrielides&Sukow}, owing to the simplicity of implementation and feasibility of extension to high-speed operation. 
Delayed-feedback laser systems also have the potential to generate very high-dimensional and complex chaotic dynamics \cite{ikeda-prl} and strategies for controlling fast chaos exist for these type of systems \cite{control_paper}.

In this paper we describe a new fast optical time-delay feedback device with flexible dynamical timescales and complexity. Adjusting the pass-band characteristics of the feedback loop allows tuning of the characteristic time scale while the time-delay and the feedback strength control the complexity of the dynamics.
 The device belongs to the class of optical systems with passive nonlinearity in the feedback loop (see Fig.~\ref{fig:experimental-setup}). 
 In our device a Mach-Zehnder interferometer is the source of nonlinearity while the semiconductor laser that provides the optical power acts as a linear current-to-optical-frequency converter.
The nonlinearity of the interferometer coupled with the delay in the feedback loop combine to
produce a range of steady state, periodic, and chaotic behavior.

The flexible time scale allows us to operate the device at moderate speeds to perform detailed system characterization while it can be also operated at high-speed for applications like chaos communication and control of fast chaos.
In this paper we tune the device to a moderate speed so that it generates dynamics with frequencies of hundreds of MHz.
Another advantage of the device is that the nonlinearity (interferometer) is easily accessible and tunable so that it can be reproduced and controlled accurately, without involving the internal dynamics of semiconductor lasers as in other devices.
Furthermore, this system is constructed with relatively
inexpensive components making it an economical choice as a chaotic optical
source in future applications.

A distinguishing feature of our device is that it uses an AC-coupled amplifier in the feedback loop so that it has bandpass characteristics. It can thus operate, in principle, in the radio frequency or microwave range using readily available components. This contrasts previous research that used DC-coupled low-pass filter components \cite{ohtsubo-optics-comm91,ohtsubo-optics-comm92,ohtsubo-qe}. Time-delay dynamical systems with bandpass filtering has only recently been investigated and has the advantage that the bandwidth of the chaotic signal can be tailored to fit the desired communication band \cite{goedgebuer-qe,goedgebuer-cs}.

The goal of this paper is to present details about the experimental implementation of our new device and to develop a model that allows us to investigate the nonlinear dynamics of the system. A thorough characterization of the system and a good model are prerequisites for investigating applications such as control of fast chaos, that will be reported elsewhere \cite{control_paper}. 
We describe the experiment in Sec. \ref{section:experimental_setup} and develop a deterministic model for the device in Sec. \ref{section:model}. Subsequently, we investigate in Sec. \ref{section:hopf} the Andronov-Hopf bifurcation of the steady state. Finally, we will present evidence that our opto-electronic feedback system generates high-dimensional chaos in Sec. \ref{section:chaos}.

\section{Experimental setup}
\label{section:experimental_setup}

\begin{figure}
\centerline{\includegraphics*[width=0.8 \columnwidth]{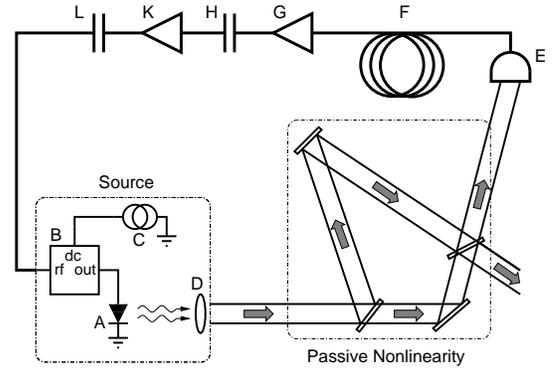} }
\caption{Schematic of experimental setup. The device consists of a voltage controlled source, a passive nonlinearity, and a feedback loop with bandpass characteristics. The components labeled A-L and details of the setup are explained in the text. }
\label{fig:experimental-setup}
\end{figure}

In this section we describe details of the experimental implementation of the 
active laser interferometer with AC-coupled feedback. The device consists of the laser, that acts a current controlled source, the interferometer, that constitutes the passive nonlinearity in the system, and the feedback loop with bandpass characteristics. A schematic of the experimental setup is shown in Fig.~\ref{fig:experimental-setup} where the labels A-L correspond to components that we refer to and describe below.

The light source is an AlGaInP diode laser (A - Hitachi HL6501MG, wavelength 0.65 $\mu m$) with a multi-quantum well
structure. The diode is housed in a commercial mount (B - Thorlabs TCLDM9)
equipped with a bias-T for adding an RF component to the injection current.
 Thermoelectric coolers in the mount are connected to a
proportional-integral-derivative feedback controller (Thorlabs TEC2000) to 
provide $1$ mK temperature stability thereby minimizing frequency and power drift. The output light of the laser is collimated by a lens 
(D - Thorlabs C230TM-B, f=4.5mm) producing an elliptical beam (1 mm $\times 
$ 5 mm) with a maximum output power of 35 mW.

The passive nonlinearity in the experiment consists of a Mach-Zehnder interferometer with unequal path lengths (path difference 45 cm) into which the laser beam is directed. A silicon photodetector (E - Hammamatsu S4751, DC-750
MHz bandwidth, 15 V reverse bias) measures the intensity of light emitted from one output port of the interferometer. 
The size of the photodiode is much smaller than the width of the
laser beam so only a fraction of the interferometer's output is detected.
The small detector size ensures that only one fringe appears within the beam cross section thus compensating for wavefront aberrations and slight laser beam misalignment and improving the fringe visibility.
A neutral density filter is
fixed to the front of the laser mount limiting the optical power reaching
the photodiode to prevent saturation.

The feedback-loop photodiode produces a current proportional to the
optical power falling on its surface. The current flows through a $50\Omega $
resistor. The voltage across that resistor is transmitted down a coaxial
cable (F - RU 58, total length $\sim 327$ cm). 
The signal emanating from the cable passes through a low-noise, fixed-gain, AC-coupled amplifier (G - MiniCircuits ZFL-1000LN, bandwidth 0.1-1000 MHz), a DC-blocking chip capacitor (H - 220 pF), an AC-coupled amplifier (K - Mini-Circuits
ZFL-1000GH, bandwidth 10-1200 MHz), and a second DC-blocking chip capacitor (L - 470 pF). The capacitors
reduce the loop gain at frequencies below $\sim 7$ MHz where a thermal effect
enhances the laser's sensitivity to frequency modulation~\cite{Kobayashi,Tsai}.
The resulting voltage is applied to the bias-T (B) in the laser mount. The bias-T converts the signal into a current and
adds it to a DC injection current from a commercial laser driver (C - Thorlabs LDC500).

The entire system is fixed on an optical table using short (2 inch) mounts
for mechanical stability. This stability is extremely important as variation
in the path length on the order of the wavelength of the laser light
(0.65 $\mu $m) produces significant power variations at the output of the
interferometer. Furthermore, the entire apparatus is covered by an
insulating box to prevent thermal expansion or contraction of the mirror
mounts due to air currents.

%%%%%%%%%%%%%%%%%%%%%%%%%%%%%%%%%%%%%%%%%%%%%%%%%%%%
% FIGURE 1
%%%%%%%%%%%%%%%%%%%%%%%%%%%%%%%%%%%%%%%%%%%%%%%%%%%%
%
%
% 
%%%%%%%%%%%%%%%%%%%%%%%%%%%%%%%%%%%%%%%%%%%%%%%%%%%%

\section{Mathematic model of the opto-electronic device}
\label{section:model}

To obtain a model of the device, we consider in turn the relevant physics of the laser diode, the Mach-Zehnder interferometer, and the feedback loop components. 

\subsection{Source: Semiconductor Laser}

The injection current applied to the laser diode is a combination of the DC-bias current and the high-frequency currents due to the time-delayed output of the feedback loop. Modulating semiconductor lasers by varying the input current results primarily in changes of the laser frequency and, to a lesser extend, the laser power.
One mechanism relating the input current and frequency shifts is the change of carrier density in the laser device as result of the modulation. A changed carrier density shifts the refractive index of the material that makes up the laser cavity and thereby changes the frequency of the lasing mode.
A second mechanism that leads to changes in the optical frequency is a thermal effect that enhances frequency modulations below $\sim 5$ MHz by as much as a few orders of magnitude \cite{Kobayashi,Tsai}

If the pumping current is modulated at a rate significantly slower
than the GHz internal time scale typical for semiconductor 
lasers, then the output power and frequency will adiabatically follow the input so that they depend in a linear fashion on the current when the laser is operated far above threshold.

To determine the main internal time scale of the laser dynamics in the absence of time-delay feedback we determine its relaxation oscillation frequency, $\Omega_R$,  by  measuring the peak in the power spectrum of the intensity noise using a high-speed spectrum analyzer (Hewlett-Packard 8566B, 22 GHz bandwidth).
We find, as expected \cite{agrawal}, that  $\Omega_R^2 \propto i-i_{th}$ where $i$ is the DC-injection current and $i_{th}$ is the laser threshold injection current. The relaxation oscillation frequency at the nominal operating current of 75 mA used in the experiment is $\Omega_R=2.7$ GHz, and $i_{th}=40.7$ mA.

Based on our measurements, we see that the laser will adiabatically follow the injection current for frequencies much less than $\sim 2.5$ GHz (bounded by $\Omega_R$) and much greater than $\sim 5$ MHz (bounded by the enhanced frequency tuning due to the thermal effect). Our feedback loop is bandpass limited to within this range and hence the laser can be modeled as a voltage controlled source, a linear device which converts variations of the input voltage $U(t)$ that drives the input current into corresponding oscillations of the optical frequency $\omega(t)$ and power $P(t)$. 

The model for the laser we propose is simply
\begin{eqnarray}
P(t)  &=& \kappa \; U(t) + P_0, \label{power}\\
\omega(t) &=& \eta \; U(t) + \omega_0 = \frac{\eta}{\kappa} \left( P(t)-P_0 \right)  + \omega_0, \label{frequency}
\end{eqnarray}
where $P_0$ ($\omega_0$) denote the emission power (the optical frequency) of the steady-state, $U(t)$ is the voltage applied to the bias-T, and $\kappa$ and $\eta$ are, respectively, the voltage-to-power and voltage-to-frequency conversion strength.

We note that it is possible to obtain Eqs.~(\ref{power}) and (\ref{frequency}) from a direct analysis of the standard rate equations describing semiconductor laser dynamics \cite{agrawal} in the limit when the amplitude of the injection current modulation is small in comparison to the DC-injection current, which is appropriate for our device. For higher-frequency operation of the device, above $\Omega_R$, the full semiconductor laser rater equations would have to be used in place of  Eqs.~(\ref{power}) and (\ref{frequency}). Such high-frequency modulation can lead to nonlinear phenomena such as period-doubling cascades, period tripling, and chaos \cite{dml-exp2} when the modulation amplitude is large. We do not consider further such behavior.

\subsection{Passive Nonlinearity: Mach-Zehnder Interferometer}

The passive nonlinearity in our system is an unequal-path Mach-Zehnder interferometer that converts frequency variations of the light emitted by the laser into intensity variations at the photodetector. The photocurrent produced by the detector is converted into a voltage $V_{det}(t)$ by a resistance $r=$ 50 $\Omega$ and is subsequently amplified. The difference in propagation time between the two paths of the interferometer is $\Delta_t=1.5$ ns, much smaller than the time scale of variation of the laser injection current. 
 Therefore, we can assume that the light waves that reach the detector through the two different paths have the same optical frequency. The voltage $V_{det}$ is thus given by
\begin{equation}
V_{\det }(t)=\gamma~P(t)\left\{ 1+b~\cos \left[\omega(t) \, \Delta_t \right] \right\},  \label{detector_voltage_long}
\end{equation}
where $\omega(t)$ is the optical frequency. 
The parameter $\gamma$ describes the overall feedback strength which is determined by the fraction of the power in the laser beam that actually falls onto the detector, the sensitivity of the photodiode, and the electronic amplification of the signal.
The parameter $b \in [0,1]$ is the fringe visibility and is defined as
\begin{equation}
b = \frac{P_{max}-P_{min}}{P_{max}+P_{min}},
\end{equation}
 $b=1$ corresponds to the ideal situation where the light waves in the interferometer are
perfect plane waves and where the beamsplitters divide them exactly in half.
In our experiment we find $b \approx 0.8$.  

The optical frequency is related to the observed output power by Eq.~(\ref{frequency}), which allows us to rewrite Eq.~(\ref{detector_voltage_long}) as
\begin{equation}
V_{\det }(t)=\gamma~P(t)\left\{ 1+b~\sin \left[ \alpha \, \left(P(t)-P_0\right) + \phi \right] \right\},  \label{detector_voltage}
\end{equation}
Here, the parameter $\alpha=\eta \Delta_t / \kappa$ and  the constant offset $\phi=\omega_0 \Delta_t - \pi / 2$ are tunable. By varying the DC component of the injection current, we adjust $\phi$ to $\approx 0$. The constant parameter $\alpha$ determines the sensitivity of the interferometer and can be tuned by varying the path imbalance $\Delta_t$. 

The useful range of path-length-differences is limited by the phase noise of the laser light, which is a consequence of the quantum process of spontaneous emission. Increasing the sensitivity of the interferometer by increasing the path difference proportionally increases the effect of phase noise. We chose to set the sensitivity to a value such that the amplitude of noise fluctuations is less than 10\% of $P_0$, corresponding to $\alpha \sim$ 1.9 mW$^{-1}$.

\subsection{Electronic Feedback Loop }
%\label{sec:loop}

 %%%%%%%%%%%%%%%%%%%%%%%%%%%%%%%%%%%%%%%%%%%%%%%%%%%%
% FIGURE 2
%%%%%%%%%%%%%%%%%%%%%%%%%%%%%%%%%%%%%%%%%%%%%%%%%%%%
%
%
% 
\begin{figure}
\centerline{\includegraphics*[width=0.8 \columnwidth]{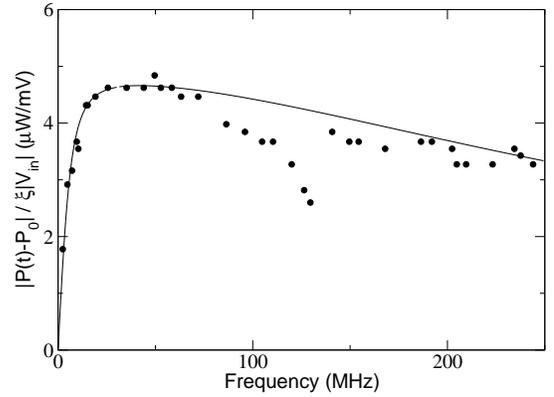} }
\caption{ Frequency response of the open loop electronic feedback for a gain value $\gamma=14.6$ mV/mW. Circles denote experimental results and the line shows the best fit of the theoretical model.}
\label{fig:openloop}
\end{figure}
%%%%%%%%%%%%%%%%%%%%%%%%%%%%%%%%%%%%%%%%%%%%%%%%%%%%

The electronic feedback loop connecting the detector and diode laser introduces a time-delay.
The total roundtrip delay-time $\tau$ is somewhat larger than this time-delay, because $\tau$ includes, for instance, the free-space propagation time of the laser light.
 However, for the purpose of modeling, we may assume all processes to be instantaneous and assign the total roundtrip delay-time to the propagation of the signal through the feedback loop.  
The frequency-limiting effects of the different components in the feedback loop are modeled by a combination of single-pole low- and high-pass filters. The feedback loop is described by 
\begin{eqnarray}
\tau_l \dot V(t) &=& - V(t) + V_{det}(t-\tau), \label{lowpass} \\
\dot U(t) &=& - \frac{U(t)}{\tau_h} + \dot V(t),   \label{highpass}
\end{eqnarray}  
where $\tau_l$ ($\tau_h$) is the low-pass (high-pass) filter time constant, $V_{det}$ is the voltage output of the photodiode, and $V$ and $U$ are the voltages at the output of the low- and high-pass filters, respectively.
The voltage $U(t)$ is used to generate the time-dependent current that is injected into the laser diode. 
Low-pass filtering in the experiment is provided by the limited bandwidth of the photodiode and the electrical connections to the laser. High-pass filtering is due to the two capacitors and the bias-T (see Fig.~\ref{fig:experimental-setup}).

Using Eq.~(\ref{power}) we can rewrite Eq.~(\ref{highpass}) in terms of the laser output power $P(t)$
\begin{equation}
\dot{P}(t) = - \frac{P(t)-P_0}{\tau_h} + \kappa \dot V(t).   \label{highpass2}
\end{equation}
We determine the parameters $\tau_l$, $\tau_h$, $\kappa$  by measuring the open-loop transfer characteristic of the electronic feedback loop. 
A signal generator is connected in place of the photodetector so that a sinusoidal voltage $V_{in}$ of known amplitude and frequency is injected into the opened feedback loop and eventually into the laser.
The optical power is measured directly at the laser output.
 Figure~\ref{fig:openloop} shows the result of the experiment and the best fit of the model (Eq.~(\ref{highpass2}) and Eq.~(\ref{lowpass}) with $V_{det}$ replaced by $\xi V_{in}$, where $\xi$ is the electronic amplification) with parameter values given in Tab.~\ref{table:parameter_values}. It is seen that the theoretical curve fits to within a few percent of the experimental data everywhere except in the region between 80 and 140 MHz. The dip in the response is caused by either the bias-T or the electronics in the commercial laser mount for which we do not have a detailed circuit diagram. 

Aside from this discrepancy, the simple model of the feedback based on single-pole low- and high-pass filters fits the experiment rather well. We choose to ignore the remaining discrepancy because the model successfully reproduces most features of the observed dynamics.

\subsection{Full Model and Parameter Values}

We obtain a full description of our opto-electronic device by combining Eq.~(\ref{lowpass}) and Eq.~(\ref{highpass2}) with the expression relating the detector voltage to the laser power, Eq.~(\ref{detector_voltage}).
All parameters in this model can be measured and are displayed in Tab.\ref{table:parameter_values}. The only parameter that was not measured directly is the time-delay $\tau$. Direct measurement of this parameter is complicated by contributions of the several electronic components, each of which introduces some unknown phase lag in addition to simple propagation delay. So we use the periodic dynamics of the device to determine $\tau$ more precisely.
 In the next section we present evidence that the steady state becomes unstable through an Andronov-Hopf bifurcation. The frequency of the resulting periodic oscillation depends on the time delay $\tau$ and we use the experimentally measured frequency close to the bifurcation point to improve the estimate of $\tau$ that we obtain by measuring the propagation delay of the feedback loop.

%
% Table
%
\begin{table}
\caption{ Definition of symbols and measured values of the model parameters.}
\label{table:parameter_values}
\begin{center}
\begin{tabular}{|c|c|l|}
\hline
Symbol&Value&Description \\
\hline
$\tau_l$ & 0.66 $\pm$ 0.05 ns		& Low-pass filter time constant\\
$\tau_h$ & 22 $\pm$ 0.5 ns		& High-pass filter time constant\\
$\tau$ 	 & 19.1 $\pm$ 0.1 ns		& Feedback delay time\\
$\kappa$ & (4.8 $\pm$ 0.1) $\frac{\mu \text{W}}{\text{mV}}$ & Modulation sensitivity\\
$\alpha$ & 1.89 $\pm$ 0.05 mW$^{-1}$   	&  Interferometer sensitivity\\
$b$      & 0.8 $\pm$ 0.02              	&  Fringe visibility \\
$P_0$    & 26 $\pm$ 0.5 mW  		&  Operating point optical power\\
$\gamma$ & 0 - 18 mV/mW 	        & Feedback gain \\
\hline
\end{tabular}
\end{center}
\end{table}

\section{First Instability: Andronov-Hopf Bifurcation}
\label{section:hopf}

%\subsection{Experimental Results}
%%%%%%%%%%%%%%%%%%%%%%%%%%%%%%%%%%%%%%%%%%%%%%%%%%%%
% FIGURE 3
%%%%%%%%%%%%%%%%%%%%%%%%%%%%%%%%%%%%%%%%%%%%%%%%%%%%
%
%
% 
\begin{figure}
\centerline{\includegraphics*[width=0.8 \columnwidth]{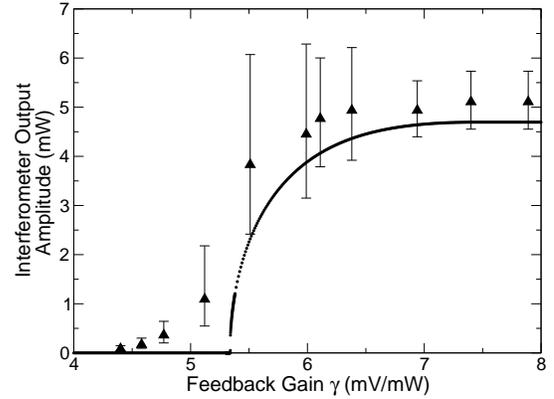} }
\caption{The oscillation amplitude measured at the second interferometer output versus gain is shown for the experiment (triangles) and model (dots).  }
\label{fig:Hopfexp}
\end{figure}
%%%%%%%%%%%%%%%%%%%%%%%%%%%%%%%%%%%%%%%%%%%%%%%%%%%%

Our opto-electronic time-delay feedback device can display very complex dynamics. As parameters of the device are changed a series of different bifurcations results in a transition from the initial steady state behavior to chaotic dynamics. In this section we discuss the first such transition in which the steady state becomes unstable and self-sustained periodic oscillations arise. 

We know of no exhaustive list of all possible ways that limit cycles (periodic oscillations) can arise in time-delay systems. However, for the well known bifurcations of cycles (those that already exist in two-dimensional systems) this list exists and by examining the scaling of the period and amplitude near the bifurcation one can distinguish between the different bifurcation scenarios \cite{book:Strogatz}. For instance, a supercritical Andronov-Hopf bifurcation is characterized by an amplitude of the stable limit cycle that scales as the square-root of the distance of the bifurcation parameter from the bifurcation point and an oscillation period of finite size that is approximately constant as the bifurcation parameter is varied. 

To investigate the bifurcations in our device, we varied the feedback gain $\gamma$, which serves as our main bifurcation parameter. A second bifurcation parameter of interest is the delay time $\tau$.
Experimentally we can change $\tau$ by adding or subtracting fixed lengths of coaxial cable to the feedback loop.

For gain values below a critical value $\gamma<\gamma_C$, the system is in a steady state with fluctuations of the observed laser output power due only to the inherent phase noise. 
When the gain is increased through the critical value $\gamma_C= 5.1 \pm 0.5$ mV/mW for $\tau \sim 19.1$ ns, the steady state is replaced by periodic oscillations. The dominant frequency of the oscillation is $51.5 \pm 1$ MHz, which is roughly equal to $1/\tau$. This frequency does not change substantially as the gain is increased further. The oscillation amplitude, on the other hand, grows smoothly from zero with increasing gain, as shown in Fig.~\ref{fig:Hopfexp}. The spontaneous emission noise of the semiconductor laser leads to an amplification of the amplitude variations (indicated by the larger error bars) close to the bifurcation \cite{Roy_PLA}. It is therefore not possible to pinpoint the bifurcation point exactly and there 
 is no clear $\sqrt{\gamma-\gamma_C}$ scaling of the amplitude, as expected for an Andronov-Hopf bifurcation.
Nevertheless, the smooth amplitude growth and the finite period of the limit cycle at $\gamma \gtrsim \gamma_C$ indicate a supercritical Andronov-Hopf bifurcation at $\gamma_C$.  In the model, which is noise-free, we find an Andronov-Hopf bifurcation at $\gamma_C=5.34$ mV/mW.

%%%%%%%%%%%%%%%%%%%%%%%%%%%%%%%%%%%%%%%%%%%%%%%%%%%%
% FIGURE 4
%%%%%%%%%%%%%%%%%%%%%%%%%%%%%%%%%%%%%%%%%%%%%%%%%%%%
%
%
% 
\begin{figure}
\centerline{\includegraphics*[width=0.8 \columnwidth]{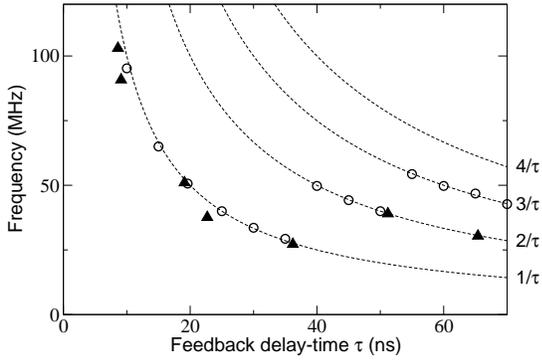} }
\caption{ Frequency of the oscillations close to the Andronov-Hopf bifurcation point versus the feedback delay time $\tau$. Measured (triangles) and numerically estimated (circles) frequencies are shown. We display curves $n/\tau$ with $n=1,2,3,4$ for visual guidance.}
\label{fig:freqvsdelay}
\end{figure}
%%%%%%%%%%%%%%%%%%%%%%%%%%%%%%%%%%%%%%%%%%%%%%%%%%%%

Next, we experimentally determined the frequency of the limit cycle close to the bifurcation point for different delay times $\tau$. In all cases the steady state becomes unstable through an Andronov-Hopf bifurcation. However,
 we find that the relation $f \sim 1/\tau$ between the frequency $f$ and the delay time $\tau$ holds only for a limited range of $\tau$. 
Figure~\ref{fig:freqvsdelay} summarizes the relation between $f$ and $\tau$ that we obtain from experimental (triangles) and numerically calculated (circles) time series. 
The data suggest that the device transitions from a steady state to limit cycle oscillations with frequencies roughly $n/\tau$, where $n=n(\tau)$ can be $1,2,3,\ldots$.

The origin of the fundamental frequency ($n=1$) can be understood intuitively by considering whether a wave circulating in the feedback loop will reinforce itself. If the feedback is positive, a wave will reinforce itself if an integer number of wavelengths equals the propagation length in the loop. If the feedback gain is negative, the propagation length must be a half integer number of wavelengths. In the experiment, we can achieve negative gain by tuning the offset $\phi$ in Eq.~(\ref{detector_voltage}) such that $\phi \approx \pi$. For this setup we observe, as expected, $f \sim 1/(2 \tau)$ for $\tau \sim 19.1$ ns (data not shown).  

The appearance of modes with $n>1$ for larger delay times $\tau$ is due to the fact that the gain in the feedback loop is not perfectly flat over the pass-band. Thus, as the gain is increased from a low level, one particular wavelength will first reach the threshold where the gain in the loop balances the losses. In a system with only low-pass feedback, the gain is highest at low frequencies, so the mode with the lowest frequency is always the one that becomes stable first independent of the delay. On the other hand, the high-pass filter introduces a bias toward high frequencies. For each mode $n$ the frequency scales as $f\sim \tau^{-1}$. This implies that the damping effect of the high pass filter on a particular mode becomes more pronounced with increasing delay time $\tau$. Finally, a higher order mode $n>1$, one that has a higher frequency for a given delay, will reach the stability threshold first.
This explains the different modes of periodic oscillations observed in the experiment. In addition, it follows from this argument that there exist specific delay times for which two modes with different frequencies have equal threshold gain $\gamma_C$ (double Hopf point). However, since none of the delay-times used in the experiment are close to one of these special $\tau$ we do not consider further this case.

\section{Chaos}
\label{section:chaos}

%%%%%%%%%%%%%%%%%%%%%%%%%%%%%%%%%%%%%%%%%%%%%%%%%%%%
% FIGURE 5
%%%%%%%%%%%%%%%%%%%%%%%%%%%%%%%%%%%%%%%%%%%%%%%%%%%%
%
%
% 
\begin{figure}
\centerline{\includegraphics*[width=\columnwidth]{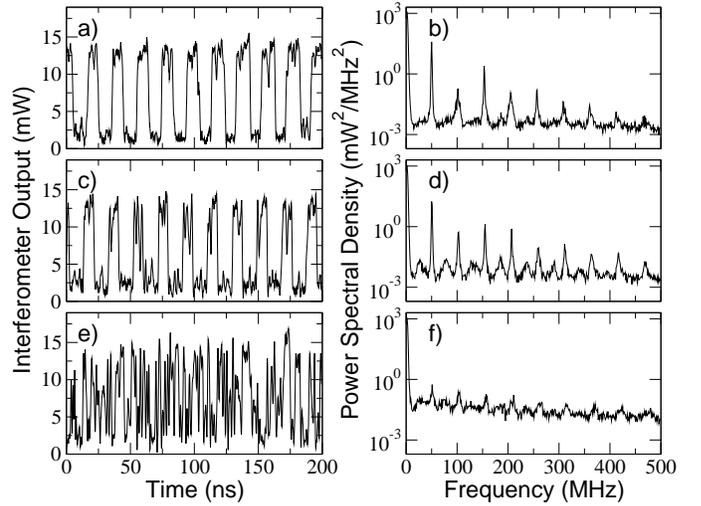} }
\caption{ Experimentally measured time series (panels a,c,e) and power spectra (panels b,d,f) obtained from the second output port of the interferometer are shown. $\gamma=9.4$ mV/mW for a) and b), $\gamma=13.2$ mV/mW for c) and d), and  $\gamma=17.6$ mV/mW for e) and f). }
\label{fig:chaos_exp}
\end{figure}
%%%%%%%%%%%%%%%%%%%%%%%%%%%%%%%%%%%%%%%%%%%%%%%%%%%%

Beyond the Hopf bifurcation successively more complex dynamics develops as the gain is increased, as shown in Figure~\ref{fig:chaos_exp}. At feedback gains higher than the Andronov-Hopf bifurcation point, the initially sinusoidal oscillations begin to square off, as shown in Fig.~\ref{fig:chaos_exp}a. The square shape of the waveform results in prominent odd harmonics in the spectrum (Fig.~\ref{fig:chaos_exp}b). As the gain increases, a small, broad peak appears at about half the fundamental frequency as shown in Fig.~\ref{fig:chaos_exp}d. The peak at roughly half the fundamental frequency is three orders of magnitude below the fundamental. The weakness and broadness of this peak coupled with the presence of phase noise may explain why no clearly period doubled behavior is apparent in the time domain (Fig.~\ref{fig:chaos_exp}c). As the gain is increased  further, the broad background rises and the tall peaks at the fundamental frequency and its harmonics weaken. The power spectrum for $\gamma=17.6$ mV/mW, shown in Fig.~\ref{fig:chaos_exp}f, is quite broad and the peaks have nearly dropped to the level of background which has risen significantly above the noise floor ($2 \times 10^{-3}$ mW$^2$/MHz$^2$ measured just below the Andronov-Hopf bifurcation point). This is indicative of high dimensional chaos in the system.

A similar very broad and featureless power spectrum  in the chaotic regime for an optical system with passive nonlinearity and bandpass feedback was reported in Ref.~\cite{goedgebuer-qe}. There, the authors synchronize two devices and successfully communicate information, thus demonstrating that the cause of the broadband spectrum is deterministic chaos.
Because of the similarity of their device to ours, we believe that the observed complex behavior for large gain values shown in Fig.~\ref{fig:chaos_exp}e is due to chaotic deterministic dynamics.

%%%%%%%%%%%%%%%%%%%%%%%%%%%%%%%%%%%%%%%%%%%%%%%%%%%%
% FIGURE 6
%%%%%%%%%%%%%%%%%%%%%%%%%%%%%%%%%%%%%%%%%%%%%%%%%%%%
%
%
% 
\begin{figure}
\centerline{\includegraphics*[width=\columnwidth]{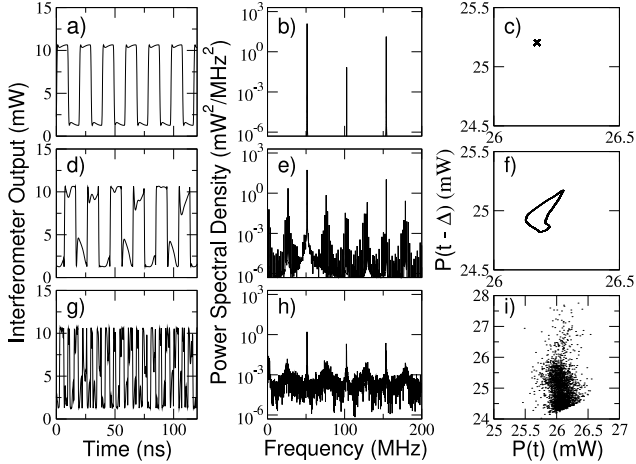} }
%\centerline{\includegraphics*[width=\columnwidth]{fig6_600dpi} }
\caption{ Numerical time series (panels a,d,g), power spectra (panels b,e,h), and Poincar\'{e} sections (panels c,f,i) are shown. 
The gain values are as in Fig.\ref{fig:chaos_exp}, that is, $\gamma=9.4$ mV/mW (a,b,c), $\gamma=13.2$ mV/mW  (d,e,f), and  $\gamma=17.6$ mV/mW (g,h,i). }
\label{fig:chaos_num}
\end{figure}
%%%%%%%%%%%%%%%%%%%%%%%%%%%%%%%%%%%%%%%%%%%%%%%%%%%%

To support this claim we show in Fig.~\ref{fig:chaos_num} time series and power spectra obtained by numerical simulation of the deterministic model\footnote{Time series from the model are obtained using an Adams-Bashforth-Moulton predictor-corrector algorithm.}. 
The match with the experimental data is good, as can be seen by comparing Fig.~\ref{fig:chaos_exp} to Fig.~\ref{fig:chaos_num}, despite the simplicity of the model and the uncertainty in the estimation of the parameters.  Figure~\ref{fig:chaos_num} also shows Poincar\'{e} sections obtained by recording the location where the trajectory uni-directionally crosses the plane $V(t)=\gamma P_0$ in the three-dimensional space spanned by $(V(t),P(t),P(t-\Delta))$ with $\Delta < \tau$. 
The simulations confirm that the system is on a limit cycle for $\gamma=9.4$ mV/mW, which is clear from the power spectrum (Fig.~\ref{fig:chaos_num}b)  and immediately obvious in the Poincar\'{e} section (Fig.~\ref{fig:chaos_num}c).
They also show that the limit cycle bifurcates to a torus-attractor for increased gain ($\gamma=13.2$ mV/mW) appearing as closed curve in the  Poincar\'{e} section (Fig.~\ref{fig:chaos_num}f). The power spectrum (Fig.~\ref{fig:chaos_num}e) exhibits a comb-like structure due to the two incommensurate frequencies of the quasi-periodic oscillation. Note, that there is not only a strong peak at $\sim$26.6 MHz, roughly half the fundamental frequency, but a definite peak at 1.8 MHz. This is well below the 3 dB cutoff point of the high-pass filter. At present we do not understand the origin of this low-frequency feature and cannot explain why the main new frequency component should be so close to one half the fundamental frequency.
For even larger gains the system is chaotic, with a very broadband spectrum (Fig.~\ref{fig:chaos_num}h) and no discernible structure in the Poincar\'{e} section (Fig.~\ref{fig:chaos_num}i). 
We computed the spectrum of Lyapunov exponents for the model with $\gamma=17.6$ mV/mW and find that the largest few exponents are positive (details on the computation of the Lyapunov spectrum can be found in Ref.~\cite{illing-qe}). This proves that the model dynamics shown in Fig.~\ref{fig:chaos_num}g are on a chaotic attractor. We estimate the attractor's Lyapunov dimension to be $D_L \sim 22$. 

In this section we have presented evidence that our device exhibits high dimensional chaos. We showed that the device undergoes a sequence of bifurcations from steady state to aperiodic oscillations with a broad and featureless power spectrum. We were able to reproduce this behavior with the deterministic model and we quantified the dimensionality of the chaotic attractor.

\section{Summary and Discussion}

We describe an optical feedback device that can produce high-dimensional chaos and that allows adjustment of the characteristic time scales of the oscillations by changing the bandpass characteristics in the feedback loop. The nonlinearity in the device is accessible and reproducible. 
 We develop a simple model that allows quantitative predictions about the behavior of the physical device and use it to determine the critical gain and frequency of the Andronov-Hopf bifurcation of the steady state. 
We observe that the device transitions to chaos with a very broadband frequency spectrum and find that this matches the model behavior.

We find that the inclusion of a high-pass filter significantly changes the qualitative dynamics of optical feedback systems with passive nonlinearity in comparison to only low-pass filtering as in the Ikeda system \cite{ikeda-prl}.
Bandpass feedback allows not only  ``fundamental'' frequencies $f \sim (2 \tau)^{-1}$ but oscillations with $f \sim \tau^{-1}$ become possible. 
The route to chaos is apparently changed when the feedback of DC-signals is blocked. That is, we do not observe a period doubling route to chaos but a more complicated transition, the details of which are not yet fully understood.

This chaotic opto-electronic device is ideally suited for both experimental investigation of fast nonlinear dynamics and technological application of high-speed chaos. For example, we use it to investigate control of fast chaos and  
are able to successfully stabilize a periodic orbit with a period of 12 ns, faster than any reported \cite{corron}. This work will be described in a later paper \cite{control_paper}. 
Also, by adjusting the time-scale of the oscillation the device could be made to oscillate at GHz frequencies suitable for use in a practical chaos communication system.


\begin{thebibliography}{99}

%
% Optical Communications with delay systems
%
% 1 laser systems 
%
%  Experiment
%
\bibitem{greg&roy-science} G. D. VanWiggeren, R. Roy, ``Communication with chaotic lasers,'' \emph{Science}, vol. 279, pp. 1198-200, 1998
% no.5354, 20 Feb. 1998, 

\bibitem{goedgebuer-prl}  J. P. Goedgebuer, L. Larger, H. Porte, ``Optical cryptosystem based on synchronization of hyperchaos generated by a delayed feedback tunable laser diode,'' \emph{Phys. Rev. Lett.}, vol. 80, pp. 2249-52, 1998 %no.10,pp.2249-2252, 9 March 1998


\bibitem{greg&roy-prl1} G. D. VanWiggeren, R. Roy, ``Optical communication with chaotic waveforms,'' \emph{Phys. Rev. Lett.}, vol. 81, pp. 3547-50, 1998
% no.16, 19 Oct. 1998, pp.3547-50. Publisher: APS, USA.

\bibitem{Davis_pra} I. Fischer, Y. Liu, P. Davis, ``Synchronization of chaotic semiconductor laser dynamics on subnanosecond time scales and its potential for chaos communication,'' \emph{ Phys. Rev. A}, vol. 62, pp. 011801, 2000 %June 2000

\bibitem{Ohtsubo_opt_lett} K. Kusumoto, J. Ohtsubo, ``1.5 GHz message transmission based on synchronization of chaos in semiconductor lasers,'' \emph{Opt. Lett.}, vol. 27, pp. 989-91, 2002
%Optics Lett., Vol. 27, No.12, pp. 989-991, June 2002

\bibitem{greg&roy-prl2} G. D. VanWiggeren, R. Roy, ``Communication with dynamically fluctuating states of light polarization,'' \emph{Phys. Rev. Lett.}, vol. 88, pp. 097903, 2002
%Physical Review Letters, vol.88, no.9, 4 March 2002, pp.097903/1-4. Publisher: APS, USA. 

\bibitem{illing_prl}  N. F. Rulkov, M. A. Vorontsov, L. Illing, ``Chaotic Free-Space Laser Communication over a Turbulent Channel,'' \emph{Phys. Rev. Lett.}, vol. 89, pp. 277905, 2002
%Phys. Rev. Lett. 89, 277905, Dec. 2002 



%
% Control
%
\bibitem{Pyragas} A. Namaj\={u}nas, K. Pyragas, A. Tama\v{s}evi\v{c}ius, ``Analog techniques for modeling and controlling the Mackey-Glass system,'' \emph{Int. J. Bif. Chaos}, vol. 7, pp. 957-62, 1997

\bibitem{Meucci} A. Labate, M. Ciofini, R. Meucci,  ``Controlling quasiperiodicity in a CO$_2$ laser with delayed feedback,'' \emph{Phys. Rev. E}, vol. 57, pp. 5230-36, 1998

\bibitem{Gavrielides} A. Hohl, A. Gavrielides, ``Experimental control of a chaotic semiconductor laser,'' \emph{Opt. Lett.}, vol. 23, pp. 1606-08, 1998
%No. 20, Octobrt 15 1998

\bibitem{Gauthier&Sukow} D. W. Sukow, D. J. Gauthier, ``Entraining power-dropout events in an external-cavity semiconductor laser using weak modulation of the injection current,'' \emph{IEEE J. Quantum Electron.}, vol. 37, pp. 175-83, 2000
%vol.36, no. 2, Feb 2000, Page(s): 175-183
 
\bibitem{Gavrielides&Sukow} F. Rogister, D. W. Sukow, A. Gavrielides, P. Mgret, O. Deparis, M. Blondel, ``Experimental demonstration of suppression of low-frequency fluctuations and stabilization of an external-cavity laser diode,'' \emph{Opt. Lett.}, vol. 25, pp. 808-10, 2000
%Optics Letters, Volume 25, Issue 11, 808-810, June 2000 


%%%%%%%%%%%%%
%
%        Ikeda 
%
\bibitem{ikeda-prl} K. Ikeda, K. Kondo, ``Successive higher-harmonic bifurcations in systems with delayed feedback,'' \emph{Phys. Rev. Lett.}, vol. 49, pp. 1467-70, 1982
% vol.49, no.20, 15 Nov. 1982, pp.1467-70. 

%\bibitem{ikeda-physicaD} Kensuke Ikeda, Kenji Matsumoto, High-dimensional chaotic behavior in systems with time-delayed feedback, Physica D: Nonlinear Phenomena, Volume 29, Issues 1-2, November-December 1987, Pages 223-235.




%
% Control Paper to be submitted
%
\bibitem{control_paper} J. N. Blakely, L. Illing, Daniel J. Gauthier, ``Controlling Fast Chaos'', in preparation 




%
% Ohtsubo
%

%\bibitem{ohtsubo-optics-lett} J. Ohtsubo, Y. Liu, ``Optical bistability and multistability in an active interferometer,'' \emph{Opt. Lett.}, vol. 15, pp. 731-3, 1990
%. [Journal Paper] Optics Letters, vol.15, no.13, 1 July 1990, pp.731-3. USA. 

\bibitem{ohtsubo-optics-comm91} Y. Liu, J. Ohtsubo, ``Observation of higher harmonic bifurcations in a chaotic system using a laser diode active interferometer,'' \emph{Opt. Comm.}, vol. 85, pp. 457-61, 1991
%. [Journal Paper] Optics Communications, vol.85, no.5-6, 1 Oct. 1991, pp.457-61. Netherlands. 

\bibitem{ohtsubo-optics-comm92} Y. Liu, J. Ohtsubo, ``Period-three cycle in a chaotic system using a laser diode active interferometer,''  \emph{Opt. Comm.}, vol. 93, pp. 311-17, 1992
%. [Journal Paper] Optics Communications, vol.93, no.5-6, 15 Oct. 1992, pp.311-17. Netherlands. 

\bibitem{ohtsubo-qe} T. Takizawa, Y. Liu, J. Ohtsubo, ``Chaos in a Feedback Fabry-Perot Interferometer'', \emph{IEEE J. Quantum Electron.}, vol. 30, pp. 334-8, 1994

%goedgebuer-pre,goedgebuer-qe
% Goedhebuer
%

%\bibitem{goedgebuer-pre} J. P. Goedgebuer, L. Larger, H. Porte, F. Delorme, ``Chaos in wavelength with feedback tunable laser diode,'' \emph{Phys. Rev. E}, vol. 57, pp. 2795-98, 1998
% Physical Review E, vol. 57, no. 3, pp. 2795-2798, March 1998

\bibitem{goedgebuer-qe} J. P. Goedgebuer, P. Levy, L. Larger, C. C. Chen, W. T. Rhodes, ``Optical Communication With Synchronized Hyperchaos Generated Electrooptically,'' \emph{IEEE J. Quantum Electron.}, vol. 38, pp. 1178-83, 2002
%, IEEE Journal of Quantum Electronics, vol. 38, no. 9, , pp. 1178-1183, September 2002

%\bibitem{goedgebuer-prl}  J.P. Goedgebuer, L Larger, and H. Porte,''Optical cryptosystem based on synchronization of hyperchaos generated by a delayed feedback tunable laser diode'', Phys.Rev.Lett., vol.80, no.10,pp.2249-2252, 9 March 1998

\bibitem{goedgebuer-cs} V. S. Udaltsov, L. Larger, J. P. Goedgebuer, M. W. Lee, E. Genin, W. T. Rhodes, ``Bandpass Chaotic Dynamics of Electronic Oscillator Operating With Delayed Nonlinear Feedback,'' \emph{IEEE Trans. Circuits Syst. I}, vol. 49, pp. 1006-09, 2002
% IEEE Trans. Circuits Syst. I, vol. 49, no. 7, pp. 1006-1009, July 2002





%%%%%%%%%%%%%%%%%%%%%%%%% Thermal Effect %%%%%%%%%%%%%%%%%%%%%%%%
%
%{Kobayashi,Tsai}

\bibitem{Kobayashi} S. Kobayashi, Y. Yamamoto, M. Ito, T. Kimura,  ``Direct frequency modulation in AlGaAs semiconductor lasers,'' \emph{IEEE J. Quantum Electron.}, vol. 18, pp. 582-95, 1982
% Volume: 18,   Issue: 4,   Year: Apr 1982 Page(s): 582- 595
 
\bibitem{Tsai}  C. Y. Tsai, C. H. Chen, T. L. Sung, C. Y. Tsai, J. M. Rorison, ``Theoretical modeling of carrier and lattice heating effects for frequency chirping in semiconductor lasers,'' \emph{Appl. Phys. Lett.}, vol. 74, pp. 917-19, 1999
%vol. 74, no. 7, pp. 917-919, 15 Feb 1999
 



\bibitem{agrawal} G. P. Agrawal, N. K. Dutta, \emph{Long-Wavelength Semiconductor Lasers}, New York: Van Nostrand Reinhold Company, 1986 


%%%%%%%%%%%%%%%%%% Directly Modulated Lasers -- Nonlinear %%%%%%%%%%%%%%%%%%%%%%%%% 



% experimental results
  
\bibitem{dml-exp2} H.~F. Liu, W.~F. Ngai, ``Nonlinear dynamics of a
  directly modulated 1.55 $\mu$m InGaAsP distributed feedback
  semiconductor laser,'' {\em IEEE J. Quantum Electron.}, vol. 29,
  pp. 1668-75, 1993 % no.6,


%%%%%%%%%%%%%%%%%%%%%%%%%%%%%%%%%%%%%%%%%%%%%%%%%%%%%%%%%%%%%%%%%%%%%%%%%%%%%%%%%%%%%%



\bibitem{book:Strogatz}  S. H. Strogatz, \emph{Nonlinear Dynamics and Chaos}, Perseus Books Publishing,  pp. 264, 1994


\bibitem{Roy_PLA} J. Garcia-Ojalvo, R. Roy, ``Noise amplification in a stochastic Ikeda model,'' \emph{Phys. Lett. A}, vol. 224, pp. 51-56, 1996

\bibitem{illing-qe} H. D. I. Abarbanel, M. B. Kennel, L. Illing, S. Tang, H. F. Chen, J. M. Liu, ``Synchronization and communication using semiconductor lasers with optoelectronic feedback,'' \emph{IEEE J. Quantum Electron.}, vol. 37, pp. 1301-11, 2001


\bibitem{corron} N. J. Corron, B. A. Hopper, S. D. Pethel, ``Limiter control of a chaotic RF transistor oscillator,'' \emph{Int. J. Bif. Chaos}, vol. 13, pp. 957-61, 2003

\end{thebibliography}
\end{document}